# Forecast-Driven Scenario Generation for Building Energy Management Using Stochastic Optimization

Hossein Nourollahi Hokmabad, *Student Member, IEEE,* Tala Hemmati Shahsavar, *Member, IEEE*, Pedro P. Vergara, *Senior Member, IEEE*, Oleksandr Husev, *Senior Member, IEEE*, and Juri Belikov, *Senior Member, IEEE*

*Abstract*—Buildings are essential components of power grids, and their energy performance directly affects overall power system operation. This paper presents a novel stochastic optimization framework for building energy management systems, aiming to enhance buildings' energy performance and facilitate their effective integration into emerging intelligent power grids. In this method, solar power generation and building electricity demand forecasts are combined with historical data, leveraging statistical characteristics to generate probability matrices and corresponding scenarios with associated probabilities. These scenarios are then used to solve the stochastic optimization problem, optimizing building energy flow while accounting for existing uncertainties. The results demonstrate that the proposed methodology effectively manages inherent uncertainties while maintaining performance and outperforming rule-based and custom build reinforcement learning based solutions.

## I. Introduction

With the integration of solar photovoltaic (PV) systems in buildings and the widespread deployment of distributed energy resources, local energy management agents are becoming crucial in the energy transition. However, these platforms face challenges in optimizing energy use due to the intermittent nature of renewables and the variability of building electricity demand [1]. Inclusion of Battery Energy Storage (BES) and Electrical Vehicles (EV) further increases system complexity by introducing additional parameters that complicate decision-making [2].

Building energy management methodologies comprise rule-based, deterministic, stochastic, Model Predictive Control (MPC), and Artificial Intelligence (AI)-based approaches, including Reinforcement Learning (RL). Rule-based methods dominate commercial applications due to their simplicity but offer limited adaptability. Deterministic approaches leverage predictions for optimization but struggle with uncertainty and rely on forecast accuracy. Stochastic methods address uncertainties in power flow optimization but require high computational effort due to scenario-based modeling.

Stochastic optimization algorithms include heuristic approaches, such as Monte Carlo methods, which use random sampling to account for uncertainty in optimization problems. These methods generate a wide range of potential solutions based on probabilistic simulations [3]. Metaheuristic algorithms, such as genetic algorithms [4] and particle swarm optimization [5], are widely adopted approaches for finding near-optimal solutions to complex optimization problems.

Real-time optimization methods are essential for the practical implementation of BEMSs. In this category, stochastic MPC-based systems provide promising results due to their ability to compensate for forecasting errors in real-time. The authors in [6], reported performance very close to ideal conditions when future system conditions are well-known. In [7], a real-time Mixed Integer Linear Programming (MILP) based, real-time stochastic BEMS is presented. In [8], authors utilized stochastic optimization to design self-healing BEMs, which is capable of handling real-time contingencies based on its available resources. Model-free and data-driven solutions are also gaining significant attention for solving power flow optimization problems. With advancements in computing power, AI- and RL-based methods are being widely applied and have demonstrated reliable performance. For instance, RL-based solutions inherently handle system uncertainties. For example, the Deep-RL-based model-free method proposed in [9], outperforms stochastic programming-based methods for power flow optimization. However, the study did not compare the computational complexity of the proposed solution, or the amount of data required for model training. Despite the progress in BEMSs, this field is still relatively new and requires further research to develop reliable solutions that can be confidently integrated into real-world platforms while ensuring performance that justifies the owner's investment.

To further investigate and enhance the performance of stochastic programming solutions, this work presents a forecast-driven approach for generating scenarios for both demand and PV power production to optimize power flow and improve BEMS performance in buildings equipped with PV systems and BES. The proposed approach utilizes historical data distributions to calculate probabilities and pairs generation, and demand profiles based on probability-aware sampling of feasible possibilities. The most probable scenarios are then selected and used to solve a multi-scenario

The work of J. Belikov and H. N. Hokmabad was partially supported by the Estonian Research Council (grant No. PRG1463), and by the Estonian Centre of Excellence in Energy Efficiency, ENER (grant TK230) funded by the Estonian Ministry of Education and Research.

H. NH. is with Tallinn University of Technology, Tallinn, Estonia (e-mail: hossein.nourollahi@ taltech.ee).
T. HS. is with ABB Ltd, Tallinn, Estonia (e-mail: tala.hemmati@gmail.com).
P. PV. is with Delft University of Technology, Delft, Netherlands (email: p.p.vergarabarrios@tudelft.nl).
O. H. is with Gdansk University of Technology, Gdansk, Poland (e-mail: oleksandr.husev@pg.edu.pl).
J. B. is with Tallinn University of Technology, Tallinn, Estonia (mail: juri.belikov@ taltech.ee).

optimization problem formulated as a MILP model. By incorporating forecasting tools in the scenario generation and selection procedure, the approach enhances system performance and provides robust solution that accounts for uncertainties.

## II. Proposed Methodology

### A. Building Energy Management Systems

Near Zero Energy Buildings' electricity networks generally consist of loads, a solar PV system, battery energy storage, and power electronics infrastructure. Power electronics tools are essential for enabling power exchange and conversion within the building network and with the electricity grid.

Building Energy Management System (BEMS) serves as a supervisory element, collecting near real-time indigenous data from the building's electricity network using IoT sensors and power electronics devices, as well as the required external data, such as weather conditions and electricity tariffs, from the internet. These data are utilized to optimize energy flow within the building to achieve predefined goals, such as minimizing energy costs or maximizing self-consumption. Figure 1 provides a schematic representation of BEMS and its components.

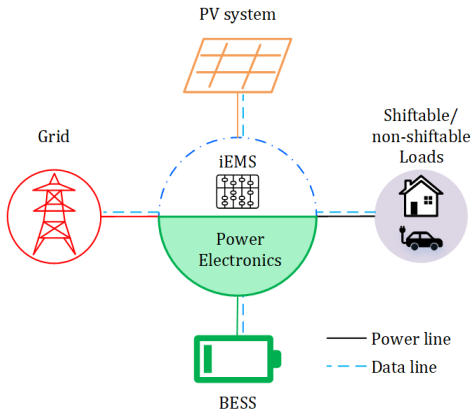

Fig. 1. Schematic representation of near zero energy buildings.

In residential buildings, loads are usually divided into two groups: shiftable and non-shiftable. Shiftable loads are those that the management system has the flexibility to reschedule based on optimization outcomes. Examples of shiftable loads include washing machines, robotic vacuum cleaners, dishwashers, and HVAC systems. Non-shiftable loads, on the other hand, are those that the BEMS cannot modify and must be served as requested. Examples of non-shiftable loads include entertainment devices, laptops, and lighting. Additionally, some loads can be categorized as hybrid loads, such as EV chargers and heat pumps, as they can function as either shiftable or non-shiftable loads depending on user requirements. In this paper, we consider all loads as non-shiftable since they are typically dominant in residential applications.

### B. Demand and Renewable Generation Forecasts

Any optimal decision-making process or an algorithm requires insight into the system's future input variables and states. Without accurate forecasting, making optimal decisions becomes highly unlikely. Conversely, if perfect forecasts were available, stochastic optimization problems could be simplified into a deterministic format. However, the intermittent and stochastic nature of electricity demand and solar PV power generation possess significant challenges to prediction accuracy.

Since discussions on forecasting methods and models are beyond the scope of this work, we have used the day-ahead solar PV power generation and demand forecasting models proposed in [10] and Long Short-Term Memory (LSTM) based Deep Neural Networks (DNN), respectively. These models generate 24 values per run, each with a 1-hour resolution for the next 24 hours. The generated values represent the expected mean values for solar PV power generation and building electricity demand or consumption at each hour of the day.

### C. Forecast-Driven Scenario Generation

As mentioned above, forecasting tools generate 24 predictions for electricity generation and consumption for the next day. Let $\bar{F}_h^G$ and $\bar{F}_h^D$ represent the forecasted values for power generation and demand at hour $h$, respectively. Assuming the forecasting models provide acceptable accuracy, these values should serve as the best possible estimates of the system's uncertain input parameters for the next 24 hours. However, the inherent uncertainties must be addressed to ensure robust decision-making.

To this end, based on available historical records and data distribution the mean and standard deviation ($\sigma$) are calculated for each hour of the day. After obtaining these values, the initial mean is replaced with the forecasted values ($\bar{F}_h^G$ and $\bar{F}_h^D$). Assuming a Gaussian distribution, a normal distribution curve is then generated using the updated mean and the previously computed $\sigma$ for each hour of the day. Since these curves represent physical quantities with finite values, they are constrained within the minimum and maximum possible ranges for electricity generation and demand.

Then, the covered range is divided into $\mathcal{R} = 100$ sections, and for each section, the probability of the actual measured value falling within that range is calculated. For instance, assuming the maximum power generation capacity of the solar PV system is 5 kWp, each section will have a resolution of 50 W. By limiting the number of sections to a fixed value, regardless of sizes of PV systems and building demand, the computational complexity remains consistent across all cases. By dividing the continuous range of possible values, we discretize and limit the potential subsequent values. However, the impact of this action is negligible in system performance.

After calculating these probabilities, a matrix of forecasting probabilities is constructed. Let $\mathcal{G}_\mathcal{P}^\mathcal{F}$ and $\mathcal{D}_\mathcal{P}^\mathcal{F}$ represent the matrices for the probabilities of potential values for electricity generation and demand, respectively, with dimensions $\mathcal{R} \times \mathcal{D}$, where $\mathcal{D}$ is the forecasting horizon, which in this case equals 24. Similarly, a probability matrix is generated for each hour of the day using previously recorded data on solar PV power generation and demand. Let $\mathcal{G}_\mathcal{P}^\mathcal{H}$ and $\mathcal{D}_\mathcal{P}^\mathcal{H}$ represent the matrices for the probabilities of historical values for generations and demand, respectively, with dimensions $\mathcal{R} \times \mathcal{D}$.

The calculation of the $D_\mathcal{P}^\mathcal{H}$ matrix is straightforward. First, recorded demand values are clustered based on their respective hours of the day and then grouped according to their corresponding power range. Once all records are classified, the probabilities for each hour and power range are computed.

However, since solar PV power generation is highly dependent on weather conditions, classifying records based solely on temporal data would lead to inaccurate results, as solar irradiance in summer is not comparable to that in winter. To address this, we propose a novel method for classifying solar PV power generation that eliminates seasonal impacts.

To achieve this, for each day of the year and each hour of the day, the maximum possible solar irradiance values are calculated based on the sun position in the sky, and building's latitude, and longitude. These values represent the theoretical maximum under clear sky conditions. The obtained solar irradiance value is then fed into a physics-based simulation of the building's solar PV system to determine the potential maximum power generation, the detailed information about physics-based modeling can be found in [10]. The minimum possible solar PV power generation value is derived from historical data by identifying the lowest recorded value for the same day and hour within a ±30-day window.

Using the obtained minimum and maximum ranges for each hour and day of the year, recorded solar PV power generation values can be categorized into a predefined number of classes. This is done by normalizing the range, determining class boundaries, and assigning each measurement to its corresponding class. In this approach, measurements are classified based on the percentage of solar PV power generation relative to the maximum feasible value. This eliminates the seasonality factor from the data, allowing for a direct comparison of solar PV power generation probabilities between summer and winter without considering the absolute magnitude of the data. Figure 2 illustrates the process described.

After assigning all measurements to their corresponding classes, the probabilities for each hour of the day are calculated for the entire dataset based on the classes that share the same hour label. As a result, the matrix $\mathcal{G}_\mathcal{P}^\mathcal{H}$ will have dimensions equal to the number of hours in a day and the number of classes, classes, which are set to 100 in this case. Then, the final probability matrices are constructed as:

$$\mathcal{G}_\mathcal{P}^\mathcal{F} + \mathcal{G}_\mathcal{P}^\mathcal{H} = \mathcal{G}_\mathcal{P}^\mathcal{T} \; , \; \mathcal{G}_\mathcal{P}^\mathcal{T} \in \mathbb{R}^{\mathcal{R} \times \mathcal{D}}, \quad (1)$$

$$D_\mathcal{P}^\mathcal{F} + D_\mathcal{P}^\mathcal{H} = D_\mathcal{P}^\mathcal{T} \; , \; D_\mathcal{P}^\mathcal{T} \in \mathbb{R}^{\mathcal{R} \times \mathcal{D}}, \quad (2)$$

where $\mathcal{G}_\mathcal{P}^\mathcal{T}$ and $D_\mathcal{P}^\mathcal{T}$, are total probabilities for production and demand ranges, respectively by considering both historical and forecasted values. Whenever probabilities are accumulated, an averaging operation is also performed to ensure that the total probability sum always remains equal to one.

Scenarios are generated based on combinations of paired power generation and consumption values. To achieve this, the Cumulative Distribution Function (CDF) for each hour of the day is derived from the $\mathcal{G}_\mathcal{P}^\mathcal{T}$ and $D_\mathcal{P}^\mathcal{T}$ probability matrices. For each CDF curve, the probability range is equally divided into $\mathcal{S}$ sections, where $\mathcal{S}$ represents number of scenarios. Then, one random value is generated for each section, and based on these values, the corresponding points are selected for each time step in the control horizon. Assuming $\mathcal{S} = 100$ for each hour of the day, there are 100 power generation ($G_h^S$) and demand ($D_h^S$) values, where $h \in [0, 23]$ and $S \in [1, \mathcal{S}]$. Finally, the daily consumption and generation profiles are generated by combining $G_h^S$ and $D_h^S$ as:

$$\mathfrak{I} = \begin{bmatrix} (G_0^1, D_0^1) & \cdots & (G_0^S, D_0^S) \\ \vdots & \ddots & \vdots \\ (G_{h-1}^1, D_{h-1}^1) & \cdots & (G_{h-1}^S, D_{h-1}^S) \end{bmatrix}_{h \times S}, \quad (3)$$

where $\mathfrak{I}$ is the matrix of all generated scenarios. Figure 3 illustrates the described procedure.

Also, since combined generation and consumption values ($G_{h-1}^S, D_{h-1}^S$) have different probabilities, the probability for each pair is considered as multiplication of each individual probability. For calculating the total probability of each scenario ($\bar{P}_s^\mathcal{T}$), the probability of each individual hour is accumulated and then averaged. These probabilities are stored in a scenario's probability matrix:

$$P_S^\mathcal{T} = [\bar{P}_1^\mathcal{T}, \bar{P}_2^\mathcal{T}, \bar{P}_3^\mathcal{T}, \ldots, \bar{P}_s^\mathcal{T}] \; , \; P_S^\mathcal{T} \in \mathbb{R}^{1 \times S}. \quad (4)$$

III. OPTIMIZATION FRAMEWORK

In the previous section, scenarios and their associated probabilities have been generated. In this section, the optimization framework will be described, to optimize building energy flow based on defined objectives.

*A. Optimization Function Formulation*

The main optimization problem consists of sub-optimization problems for each scenario. In other words, the optimal solution is the one that minimizes the defined cost function while considering all scenarios. However, this does not guarantee that the solution is optimal for each individual scenario. Thus, a general optimization problem is defined as:

$$\min_x \quad \sum_{i=1}^{\acute{s}} \mathcal{F}_i^T . X_i . \acute{P}_i^T$$

$$\text{s.t. } X_{i,\min} \leq X_i \leq X_{i,\max} \quad (5)$$

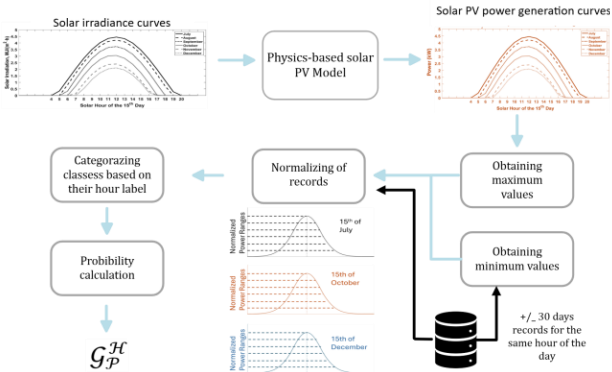

Fig. 2. Abstract representation of obtaining $\mathcal{G}_\mathcal{P}^\mathcal{H}$ from historical solar PV power generation records. Solar irradiance curves are extracted from estimation of the hourly global solar irradiation based on numerical weather predictions.

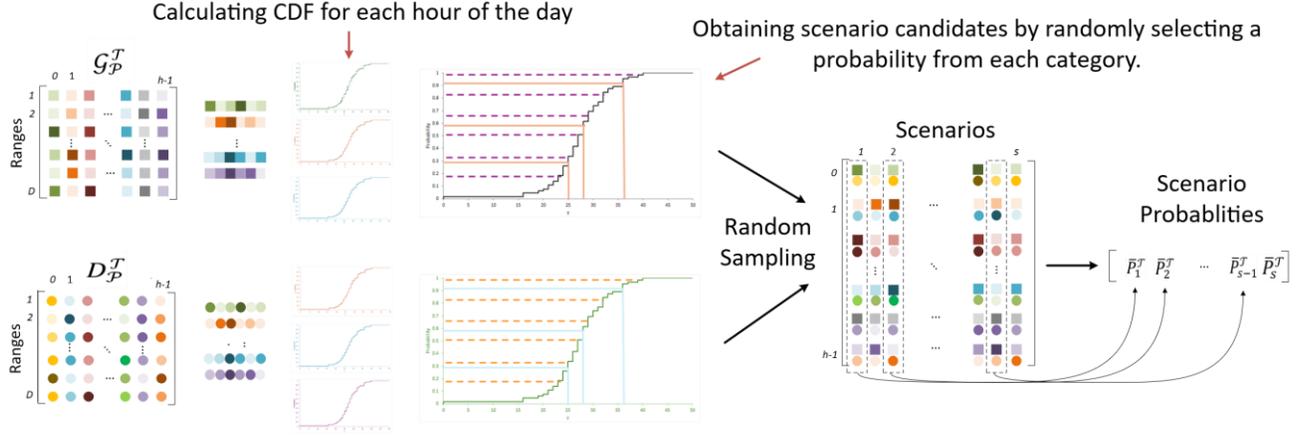

Fig. 3. Daily scenario generation based on calculated probability matrices and random candidate selection.

where $\mathcal{F}_i$ is the vector of optimization variables, $X_i$ represents the vector of optimization factors, and $\acute{P}_i^T$ denotes the probability of each scenario's occurring within the corresponding optimization horizon. Additionally, $\acute{s}$ represents the number of selected candidate scenarios. To manage the complexity of the optimization process, only the 10 most probable scenarios are selected from the generated set after sorting them by probability. It is also worth noting that the probabilities of the selected scenarios are normalized to reflect their relative differences.

Furthermore, the sub-optimization problem for each scenario is defined as:

$$\min \sum_{h=0}^{w} E_h^{imp} \times ToU_h^{imp} - E_h^{exp} \times ToU_h^{exp}. \quad (6)$$

Since energy is the time integral of power, under the assumption of constant time intervals and stable system voltage and current levels within each interval, energy parameters can be expressed as $E = P \times t$. Consequently, the optimization function can be reformulated and solved based on the power flow within the building's internal electricity network. Thus, the objectives and constraints for Eq. (6) can be represented as:

$$\forall h, i, \; 0 \leq P_{A \to B}^i[h], \quad (7)$$

$$\forall h, i, 0 \leq P_{pv \to gr}^i[h] + P_{pv \to es}^i[h] + P_{pv \to ld}^i[h] \leq P_{pv,\max}, \quad (8)$$

$$\forall h, i, \; 0 \leq P_{gr \to ld}^i[h] + P_{gr \to es}^i[h] \leq P_{gr,\max}, \quad (9)$$

$$\forall h, i, \; 0 \leq P_{es \to ld}^i[h] + P_{es \to gr}^i[h] \leq P_{es,\max}, \quad (10)$$

$$\forall h, i, P_{pv \to ld}^i[h] + P_{pv \to gr}^i[h] + P_{pv \to es}^i[h] = P_{pv}^i[h], \quad (11)$$

$$\forall h, i, P_{es \to ld}^i[h] + P_{gr \to ld}^i[h] + P_{pv \to ld}^i[h] = P_{ld}^i[h], \quad (12)$$

$$\forall h, i \; SoC_{\min} \leq SoC_{es}^i[h] \leq SoC_{\max}, \quad (13)$$

$$\forall h, i \; export_{gr}^i[h] + import_{gr}^i[h] \leq 1, \quad (14)$$

$$\forall h, i \; charge_{es}^i[h] + discharge_{es}^i[h] \leq 1, \quad (15)$$

where $E_h^{imp} = \left(P_{gr \to ld}[h] + P_{gr \to es}[h]\right) \times t$ is the total imported energy for each hour of the day from the grid, and $E_h^{exp} = \left(P_{pv \to gr}[h] + P_{es \to gr}^i[h]\right) \times t$ represents the total amount of net energy exported to the grid during hour ($h$).

Also, $ToU_h^{imp}$, and $ToU_h^{exp}$ denote the time-of-use tariffs for imported and exported energy, respectively. The notation $P_{A \to B}^i[h]$ represents power flow from point A to point B, where A, B ∈ {$pv, gr, es$}. Equation (7) ensures that all power flows are non-negative. Furthermore, $P_{pv \to gr}^i[h]$, $P_{pv \to es}^i[h]$, and $P_{pv \to ld}^i[h]$ represent power flow from PV to grid, energy storage, and load, respectively. The notation $P_{pv,\max}$ denotes the maximum allowable power output from the PV system, constrained by the PV system size and power electronics limitations. $P_{gr \to ld}^i[h]$, and $P_{gr \to es}^i[h]$, represent power flow from grid to load and energy storage, respectively, and $P_{gr,\max}$ is the maximum allowable power exchange with the grid. $P_{es \to ld}^i[h]$, and $P_{es \to gr}^i[h]$ represent power flow from energy storage to load and grid, respectively, and $P_{es,\max}$ is the maximum charge/discharge power of the energy storage unit.

Equations (11) and (12) are equality constraints ensuring that the optimization algorithm satisfies demand and utilizes all available solar PV power under all conditions. Here, $P_{pv}^i[h]$ and $P_{ld}^i[h]$ represent the generated and demanded power, at time $h$, respectively. In Eq (13), $SoC_{\min}$, $SoC_{\max}$ define, respectively, the minimum and maximum allowable battery State of Charge (SoC) levels. Finally, equations (14) and (15) prevent the optimization algorithm from generating infeasible solutions. For example, importing and exporting power to the grid simultaneously is physically impossible. Therefore, $export_{gr}^i[h]$, and $import_{gr}^i[h]$ are Boolean values, that enforce this constraint. A similar logic applies to energy storage, where $charge_{es}^i[h]$, and $discharge_{es}^i[h]$ are Boolean variables indicating the charging or discharging state of the battery. If the battery is charging $charge_{es}^i[h]$=1, and otherwise the $discharge_{es}^i[h]$=1.

### C. Performance Metrics

Various factors can be considered when evaluating the performance of EMSs. Depending on optimization goals and problem formulation, these factors may include self-consumption ratio, electricity costs, energy storage utilization and charge/discharge cycles, demand response efficiency, energy conversion losses, and more. In this study, we focus on two key performance metrics: electricity costs as the primary performance indicator and building self-consumption ratio (δ ∈ [0,100] %) as the secondary factor. However, since the optimization problem is formulated solely based on minimizing the energy bill, δ serves only as a performance measurement metric and does not influence the optimization process.

Annual energy bill ($\mathcal{A}_{bill}$) is defined as:

$$\mathcal{A}_{bill} = \sum_{d=1}^{365} \sum_{h=0}^{w} E_{h,d}^{imp} \times ToU_{h,d}^{imp} - E_{h,d}^{exp} \times ToU_{h,d}^{exp}. \quad (16)$$

And an annual self-sufficiency ratio is defined as:

$$\sigma = \sum_{d=1}^{365} \sum_{h=0}^{23} \frac{E_{pv \to ld}[h,d] + E_{pv \to es \to ld}[h,d]}{E_{pv}[h,d]}, \quad (17)$$

where $E_{pv \to ld} = P_{pv \to ld} \times t$, $E_{pv} = P_{pv} \times t$, and $E_{pv \to es \to ld} = P_{pv \to es \to ld} \times t$. The notation $E_{pv \to es \to ld}$ represents the amount of energy generated by the solar PV system, stored in the energy storage system, and later delivered to the load. This term is often overlooked in literature, where the self-consumption ratio is typically calculated only by considering the real-time power delivery from the PV system to the load.

### D. Performance Benchmarks

The proposed algorithm is benchmarked against three different approaches. The first approach is a simple rule-based algorithm that follows a priority-based energy management strategy. It prioritizes supplying demand from the PV system first, then from the BES, and finally from the grid. If the generated power exceeds the demand, the surplus energy is stored in the battery. If the battery's SoC reaches its maximum limit, any additional energy is exported to the grid. Conversely, during energy shortages, the system first utilizes stored energy in the battery, and if the demand is still not met, the remaining energy is imported from the grid.

The second approach assumes that the optimization algorithm has access to ideal forecasts for PV power generation and demand over the next 24 hours. This scenario represents the best feasible solution for an optimization problem, as it eliminates performance losses due to uncertainty, given that all future information is known and predictable. The third approach involves comparing the results with a deterministic approach, where forecasts are directly used to solve optimization problems. In this case, the forecasted values are treated as fixed, and the optimization problem is solved without considering the inherent uncertainty or variability in the system. Finally, two RL-based agents, one using Proximal Policy Optimization (PPO) and the other based on Deep Q-Networks (DQN), are utilized for benchmarking.

In all scenarios, the system follows a consistent strategy. At the start of the day, optimal control signals are generated based on forecasts for PV power generation and demand. Throughout the day, a high-resolution algorithm ensures that voltage and current levels remain within acceptable ranges. If the system falls short (e.g., PV power generation is insufficient), the grid is used to compensate for energy shortages, ensuring continuous operation within safe parameters. And if the PV generation exceeds the demand, the extra generation will be first directed to charge the BES and then the main grid if the BES SoC level reaches $SoC_{max}$.

## IV. NUMERIC RESULTS

### A. Case Study

Four years of historical consumption data from the House Number 1, obtained from the UK-DALE dataset [11] and recorded from March 2013 to April 2017, are combined with synthetic solar PV generation data for a similar period and geographical location, generated using the Photovoltaic Geographical Information System (PVGIS) system [12]. The generated dataset's granularity is 1 hour, and the recorded values are the average recorded measurement during each hour of the day. Additionally, Time-of-Use (ToU) electricity price data, sourced from Estonia's end-user electricity price dataset and recorded from March 2020 to April 2024, are utilized.

This optimization problem is formulated as a MILP model. The problem has been implemented using the Pyomo optimization framework and solved using the IPOPT solver. Additionally, all code has been developed using Python 3.11. Table I collects systems parameters, and constraints.

### B. Probability Matrices

Figure 4 (a) illustrates the number of members for distinct class labels, each corresponding to a specific range that covers the entire span of possible solar PV power generation and demand values. As shown in the figure, the probability of higher solar PV power generation is notably higher during midday hours, which is consistent across different seasons. This reflects the natural behavior of solar power generation, with peak production occurring when the sun is at its highest in the sky.

Regarding the demand patterns, which are depicted in Figure 4 (b), the probability distribution is spread across the day, but with notable peaks during the early morning hours and evening times. These peaks align with typical electricity demand patterns, as buildings typically require more power in the morning for activities such as heating, cooling, and appliance use, and again in the evening when residents return home and start using electricity for lighting, cooking, and other household tasks.

### C. Generated Scenarios

Scenarios are generated by randomly selecting and bounding consumption and generation power values for each hour of the day based on initial day-ahead forecasts. Figure 5 shows the predicted PV power generation and demand for 72 hours during the system's operation in the first week of June 2015. Observe that the PV power generation prediction outperforms the power consumption predictions. This can be attributed to the higher randomness and uncertainty associated with building power demand, whereas PV power generation is highly correlated with weather conditions and predictions. Numerical weather predictions are easily accessible from weather service providers, leading to more reliable PV power generation forecasts.

TABLE I. BUILDING ELECTRICITY NETWORK CHARACTERISTICS AND OPERATIONAL CONSTRAINTS

| Variable Name | Value | Unit | Symbol |
| --- | --- | --- | --- |
| PV system | 10 | kWp | - |
| PV inverter size | 12 | kW | $P_{pv,max}$ |
| BES capacity | 10 | kWh | - |
| BES inverter size | 5 | kW | $P_{es,max}$ |
| Grid-connected converter size | 5 | kW | $P_{gr,max}$ |
| Minimum SoC | 15 | % | $SoC_{min}$ |
| Maximum SoC | 90 | % | $SoC_{max}$ |

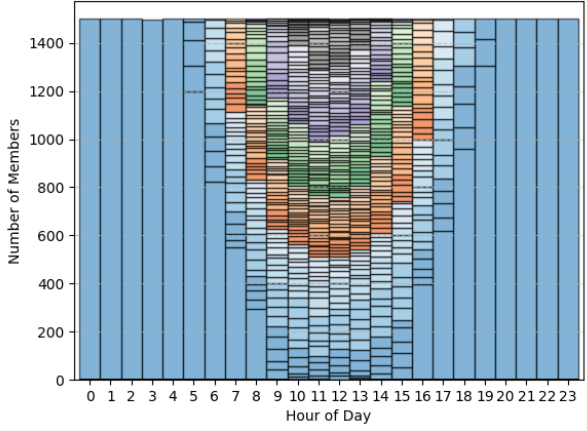

(a)

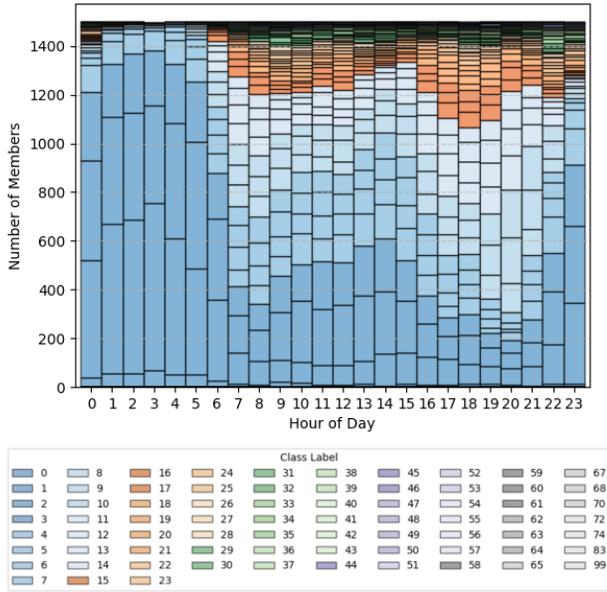

Fig. 4. (a) Solar PV system power generation class membership per hour for entire dataset. (b) Building energy demand class membership per hour for entire dataset. Each graph also represents the distribution of probabilities for each class. The classes with a higher footprint have a higher probability of occurring.

Figure 6 demonstrates the ten most probable generated scenarios for the corresponding days during the first week of June 2015. These scenarios are extracted from $\mathfrak{I}$, which is a paired combination of the $\mathcal{G}_\mathcal{P}^\mathcal{T}$ and $D_\mathcal{P}^\mathcal{T}$ probability matrices. As shown, tuning the forecasts with historical data distribution and generating scenarios based on both historical data and forecasts allowed the tool to estimate the range of PV power generation and building demand with moderate accuracy. However, improvements are still needed to ensure that the generated scenarios closely follow real profiles.

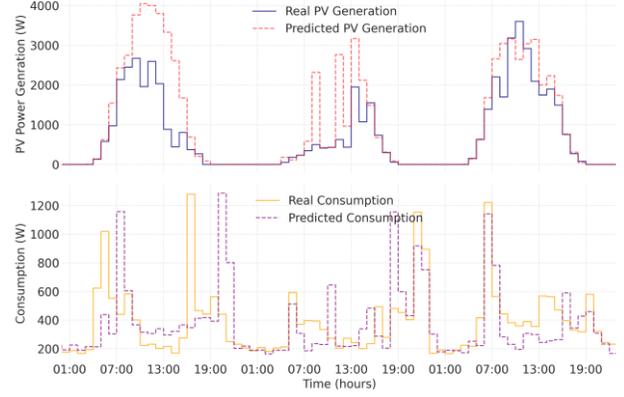

Fig. 5. Real PV power generation and building power consumption during the first week of June-2015, in comparison with their day-ahead prediction.

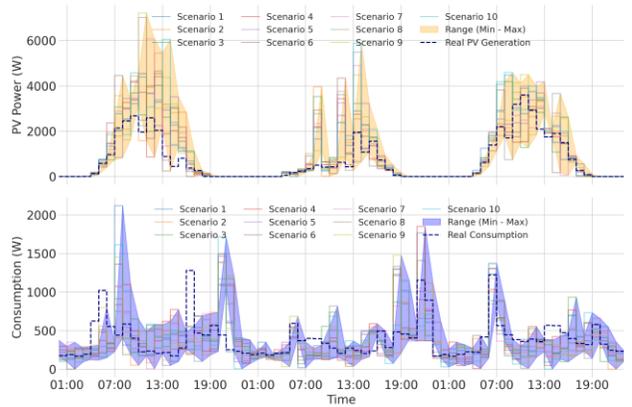

Fig. 6. Ten most probable scenarios based on day-ahead forecasts and historical data distribution.

For clarification, since PV power generation is almost zero during night hours, regardless of forecasting values or scenario outcomes, we assign zero values for these hours, , based on sun rise and set times, in the final solar PV production profiles.

### D. Performance Benchmarking

Each paired scenario is included in Eq. (5), with their corresponding probabilities as a weighting factor. In this work, we have defined the optimization problem to solely minimize the annual electricity bill; however, other factors could be incorporated to construct a multi-objective optimization problem. Figure 7 shows the optimized power flow based on input variables and scenarios. The decision made for bill minimization is highly dependent on the infrastructure size. Since the focus here is on the algorithm's performance, the results are reported considering only one configuration.

It is noticeable that the proposed methodology makes optimization decisions that are closest to the ideal forecasting scenario compared to other methodologies. However, perfect alignment is not achievable, as it is impossible to precisely forecast the system's future states. Table II presents economic and technical metrics based on one year of system operation. As expected, the ideal forecasting scenario yields the best performance both technically and economically. The closest performance is achieved by the proposed solution, which not only generates revenue but also effectively utilizes the BES to enhance the self-sufficiency ratio.

The proposed solution and the RL-DQN method exhibit relatively similar performance. However, their operational principles and computational complexities differ significantly. The primary computational burden of RL-based approaches lies in the training phase rather than inference. In contrast, the proposed method requires the calculation of probabilities and corresponding scenarios at least once per day, followed by solving an optimization problem that considers multiple scenarios. This computational demand may be regarded as a limitation of stochastic programming-based approaches. A similarity between the two approaches is their reliance on substantial historical data. Both methods face challenges related to cold-start issues, as their performance heavily depends on the availability of sufficient past data for training or scenario generation.

## V. CONCLUSION

This article presents a forecast-driven stochastic energy management agent for BEMSs. The proposed solution relies on day-ahead forecasts and historical data distributions to generate the most probable scenarios for solar PV power generation and building electricity demand. These scenarios are then used for power flow optimization to minimize the annual electricity bill while accounting for inherent system uncertainties.

The performance of the proposed method is compared with rule-based, two RL-based methods, and ideal forecasting approaches. The results confirm the superiority of the proposed method over the benchmarked models. However, it should be noted that the performance of RL-based methods heavily depends on the learning process and hyperparameters. The main disadvantages of the proposed solution are its relatively complex computational process and the need for extensive historical data, which is a typical challenge for stochastic programming-based methods.

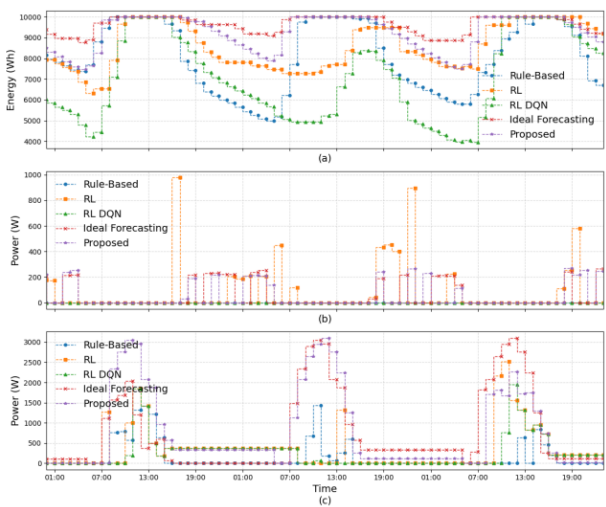

Fig. 7. The optimized power flow comparison between various methods. (a) Hourly-averaged BES stored energy levels, (b) hourly-averaged imported power from the grid, and (c) averaged exported power to the grid.

TABLE II. PERFORMANCE BENCHMARKS

| Optimization Method | SFR* (%) | AEB* (€) | ABCL* (%) | TIEG* (kWh) | TEEG* (kWh) |
|---|---|---|---|---|---|
| Rule-based | 82.01 | 135.67 | 49.38 | 2990.50 | 5451.72 |
| RL-based (PPO) | 67.02 | 5.17 | 65.64 | 5615.76 | 6426.80 |
| RL-based (DQN) | 81.48 | -70.73 | 48.94 | 3078.95 | 9384.62 |
| Ideal forecasting | 91.44 | -193.85 | 71.26 | 746.82 | 3224.71 |
| Proposed | 81.10 | -97.48 | 44.08 | 2721.38 | 4683.14 |

*SFR: PV self-consumption ratio, AEB: annual electricity bill, ABCL: average BES SoC (%), TIEG: Total imported energy from the grid, TEEG: Total exported energy to the grid.